# Differentiation of neural-type cells on multi-scale ordered collagen-silica bionanocomposites

Nicolas Debons,[†a] Dounia Dems,[†a] Christophe Hélary,[a] Sylvain Le Grill,[a] Lise Picaut,[a,b] Flore Renaud,[c] Nicolas Delsuc,[d] Marie-Claire Schanne-Klein,[e] Thibaud Coradin[a] and Carole Aimé[a,f]*

Cells respond to biophysical and biochemical signals. We developed a composite filament from collagen and silica particles modified to interact with collagen and/or present a laminin epitope (IKVAV) crucial for cell-matrix adhesion and signal transduction. This combines scaffolding and signaling and shows that local tuning of collagen organization enhances cell differentiation.

Injuries in the peripheral nervous system (PNS) are mostly caused by trauma like traffic accidents, bone fractures and joint dislocations. These injuries often lead to partial or complete loss of sensory, motor or autonomic functions that can seriously compromise the life quality of patients.[1] The PNS has a great potential for self-regeneration. However, the success of nerve self-repair crucially depends on the length of the gap. Nerve cells are able to easily bridge gaps of less than 6 mm.[2] For larger gaps, surgical grafts are used as gold standards.

Biomaterials can be used to connect damaged nerves, providing a direct framework for nerve regeneration with minimum surgery work. An optimal scaffold needs to combine the adequate anisotropic architecture to mimic aligned bundles of axons from PNS and provide guidance for neurite regrowth,[3] with suitable mechanical and surface properties to allow the formation of a new extracellular matrix (ECM) in which cells can proliferate for nerve regeneration.[4] Synthetic materials have the advantages of being easily chemically-modified and processed but can induce inflammatory reactions.[5] To avoid that, numerous studies have used materials of biological origin.[6] As major component of ECM, type I collagen is of great interest for the engineering of biomaterials and it has been processed into aligned scaffolds.[7-9]

The incorporation of bioactive molecules within anisotropic scaffolds is a topic of particular interest in tissue engineering.[10,11] From a functional point of view, of particular interest is the incorporation of laminin, a protein which is continuously synthesized after nerve injury and plays a crucial role in cell migration, differentiation and axonal growth.[12,13] A pentapeptide epitope with the IKVAV sequence (Ile-Lys-Val-Ala-Val) has been found in laminin and is known to promote neuronal differentiation and neurite outgrowth.[14,15] Composite scaffolds associating this laminin epitope with methacrylate-based polymers,[16] poly(L-lactide),[17] chitosan[18] or peptide amphiphiles[19,20] have shown that IKVAV promote neurite extension in synthetic systems. To better mimic cell microenvironment, collagen and laminin epitope (IKVAV) have been combined to investigate cell matrix adhesion and signal transduction.[21-23] However, biofunctionalization of collagen scaffold remains highly challenging since it relies on covalent conjugation. Direct conjugation of collagen triple helices before fibrillogenesis may be detrimental to their self-assembly, whereas bioconjugation after fibrillogenesis is difficult to control in terms of peptide spacing for example.[24,25] Stupp and co-workers have developed a strategy where peptide amphiphiles functionalized with the IKVAV epitope were self-assembled with collagen.[26,27] However, this approach remains challenging and time-consuming in terms of synthesis and may present limits for the amount and number of different peptides that may be loaded in the matrix. Alternatively, inorganic nanoparticles can be embedded in biopolymer networks. Inorganic particles often exhibit a rich surface chemistry and can also improve the chemical and physical stability of the biopolymer scaffold.[28] Silica nanoparticles (SiNPs) are particularly interesting candidates due to their cyto- and biocompatibility, ease of synthesis, and the versatility of sol-gel chemistry that offers various routes of biofunctionalization, including with type I collagen.[29,30] Their combination with collagen has been shown to be effective for the regeneration of bones, nerves or dermis.[31-33]

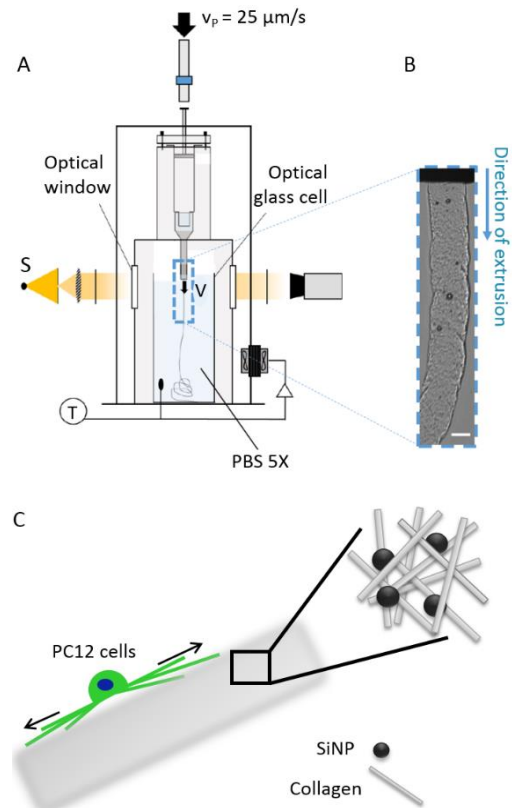

**Fig. 1.** (A) Scheme of the extrusion set-up, where *V* is the output velocity. A light source *S* is used to monitor extrusion through optical windows. *T* corresponds to the thermal regulation of the extrusion chamber. (B) Photo of an extruded filament at the needle exit in PBS 5X. Scale bar: 200 µm. (C) Scheme of a collagen-SiNP filament for PC12 adhesion and differentiation.

In this work, we engineered collagen-based composite biomaterials to improve the differentiation of PC12 cells. PC12 cells derived from the rat adrenal pheochromocytoma are widely used in neurobiology as a model for studying neuronal differentiation *in vitro*.[34-37] PC12 cells respond to fibroblast growth factor (FGF1) with robust neural processes and morphologies,[38,39] with anti-apoptotic activity.[40,41] This has been exploited for biomaterials engineering and stem cell-based therapy in neural repair.[42,43] Addition of FGF1 was used for the induction of neuronal differentiation and its quantitative analysis through the observation of cell adhesion, proliferation and differentiation (neurite number and size). Here we explored whether SiNPs can be used to improve the differentiation of PC12 cells in collagen-based biomaterials. The surface chemistry of SiNPs has been varied to impact the scaffold properties by interacting with collagen, or the signal transduction abilities by presenting the IKVAV laminin epitope. For this purpose, mono and bi-functional SiNPs were synthesized and mixed with type I collagen before extrusion of filaments. We showed that the surface chemistry of the embedded SiNPs has a key impact on their distribution within the filaments and on the collagen organization itself. Very interestingly, in our system, this scaffolding effect appeared to prevail over the expected biochemical activation from the laminin epitope.

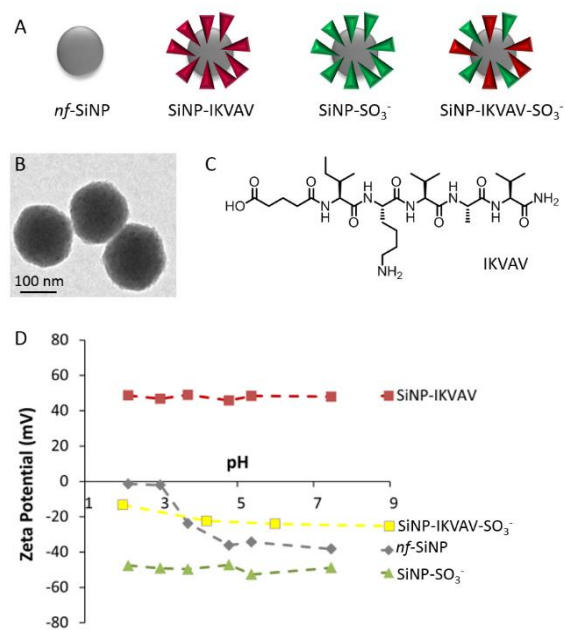

**Fig. 2.** (A) Scheme of the different SiNP surface chemistries, (B) TEM image of *nf*-SiNPs and (C) molecular structure of the IKVAV peptide. (D) Characterization of SiNPs by zeta potential measurements.

### A library of collagen-SiNP biocomposite filaments

Structuration of the scaffold is crucial to promote PN regeneration, in particular in terms of alignment. To induce the extension of neurites in the specific direction towards their synaptic target, biomaterials must stimulate their orientation growth. For this reason, bionanocomposites were extruded to produce hybrid collagen-based filaments (Fig.1). Soluble collagen and SiNPs were mixed in acidic conditions (pH 2.5). Extrusion was performed in phosphate buffered saline (PBS) 5X (pH 7.5) triggering collagen fibrillogenesis through pH increase, while ensuring the preservation of the diameter of the extruded filaments without shrinking or swelling.[44]

After having tested different collagen concentrations (13.5 and 27 mg.ml$^{-1}$) adapted to prepare soft biomaterials, and each with different SiNP concentrations ([SiNP] = 0.45; 4.5; 45 mg.ml$^{-1}$), the SiNP concentration was fixed at 45 mg.mL$^{-1}$ and collagen content to 13.5 mg.mL$^{-1}$ (Fig.S1-S4). In parallel, the surface chemistry of fluorescent SiNPs was tuned to control interactions with the collagen scaffold and/or with cells (Fig.2A). We used fluorescently labelled (Alexa 488) SiNPs (107 ± 9 nm in diameter, Fig.2B) to monitor their distribution within the filaments by fluorescence microscopy.

We have previously shown that soluble type I collagen in acetic acid interacts with sulfonate-modified SiNPs (SiNP-SO$_3^-$) by electrostatic interactions, allowing for surface-mediated collagen fibrillogenesis upon neutralization.[29] SiNPs were also modified with amine groups and conjugated with the neuroactive laminin epitope IKVAV (SiNP-IKVAV) (Fig.2C). Scaffold-interacting groups (SO$_3^-$) and cell-interacting groups (IKVAV) were also combined at the surface of SiNPs (SiNP-IKVAV-SO$_3^-$). The proper functionalization of SiNPs was checked by zeta-potential (Fig.2D). Non-functionalized SiNPs (*nf*-SiNPs) showed neutral zeta potentials at low pH and negative values from pH 4 that reflect the presence of silanol groups (pKa ca. 3.5). The presence of sulfonate groups was confirmed at low pH (below 3) where the zeta potential of SiNP-SO$_3^-$ reached - 48 mV. This is attributed to the low pKa of alkyl sulfonic acids (ca. 1). On the contrary, the zeta potentials of SiNP-IKVAV were highly positive (49 mV) in the investigated pH range, in agreement with the fact that IKVAV bears two positive charges (NH$_2$-terminal and lysine side group, pKa ca. 10). For the bifunctional SiNP-IKVAV-SO$_3^-$, zeta potentials were constant over the pH range investigated, with values ca. -10 mV in between the one of SiNP-SO$_3^-$ and of SiNP-IKVAV. This is consistent with an effective grafting of both groups at SiNP surface. The overall negative value suggests that sulfonate groups are present in a larger amount at the surface of SiNPs with respect to IKVAV.

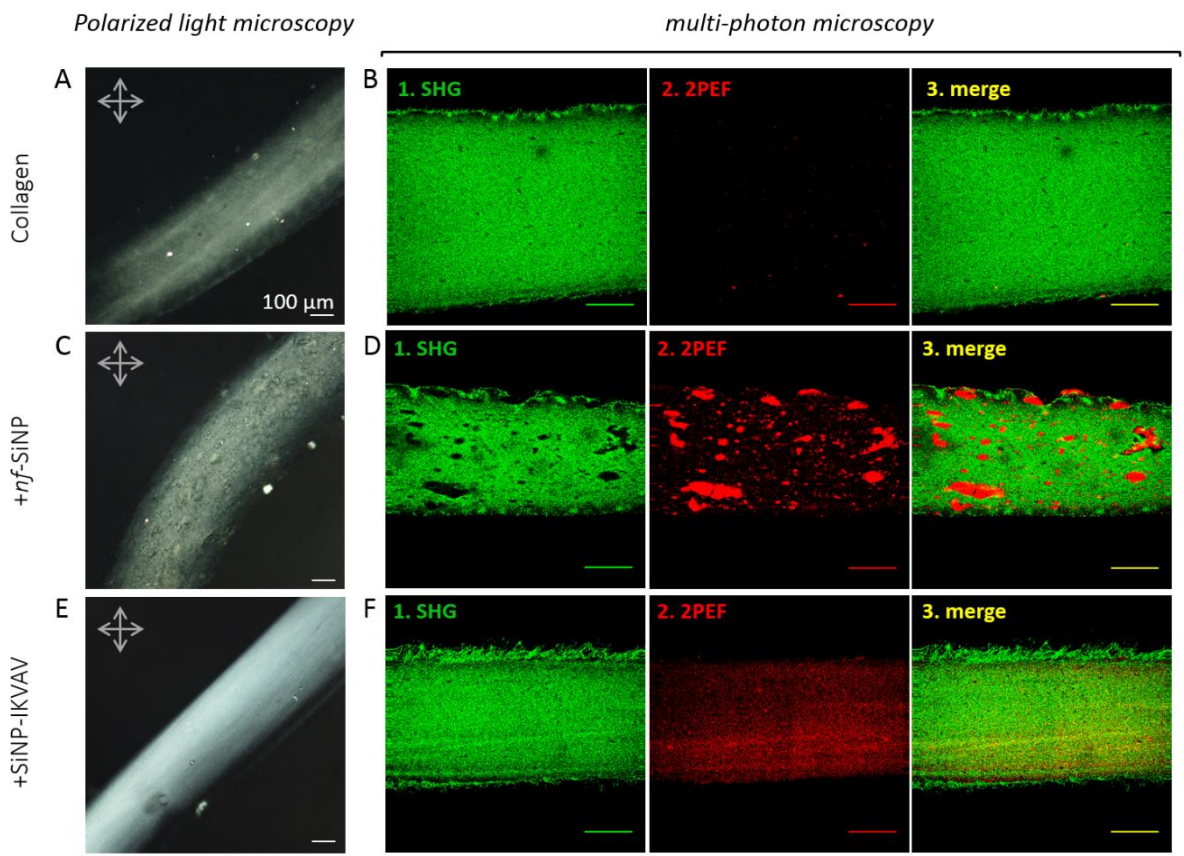

**Fig. 3.** Collagen-SiNP filaments ([Collagen] = 13.5 mg.mL$^{-1}$, [particles] = 45 mg.mL$^{-1}$) observed by (A,C,E) PLM: the uniform birefringence reflected by the brilliance of the material under PLM is indicative of uniaxial alignment (no waveplate added, arrows represent cross polarizers); and (B,D,F) multi-photon microscopy combining (1) SHG and (2) 2-PEF with (3) their merge. (A,B) Collagen; (C,D) + *nf*-SiNPs; (E,F) + SiNP-IKVAV. Scale bars 100 µm.

### Biomaterials characterization

Collagen-based filaments (ca. 300 µm in diameter) were observed under polarized light microscopy (PLM). PLM is sensitive to the birefringence of materials, in other words to optically anisotropic materials. It has proven to be effective to measure micron-scale collagen fiber orientation in tissues including tendons, ligaments, and ocular tissues.[45] It is used here to reveal the alignment of collagen molecules in the filaments. A birefringent material, with uniaxial alignment of collagen molecules, alters the polarization state of the light resulting in increase or decrease of the light intensity depending on the relative orientation between the sample and the crossed polarizers.[45] Series of images were then acquired with multiple sample orientations relative to the crossed polarizers separated by 45° (Fig. 3A,C,E, and Fig. 4A,C,E, and see also Fig.S5 to S10 for the full set of orientations). PLM observations were correlated with multiphoton microscopy combining second harmonic generation (SHG) and two-photon excited fluorescence microscopy (2PEF) modes of contrasts. SHG is a coherent process specific for non-centrosymmetric materials. At the molecular level, it relies on nonlinear dipole excitation along the peptide bonds. It builds up efficiently in highly anisotropic fibrillar collagen because collagen self-assembles into fibrils, with triple helices all oriented in the same direction with the same polarity.[46-49] This makes SHG microscopy the gold standard technique for three dimensional (3D) characterization of collagen-rich tissues with a unique structural specificity and without exogenous labelling. Simultaneously, 2PEF can be recorded to observe the distribution of fluorescent SiNPs in the filament (Fig.3B,D,F, and Fig.4B,D,F, and see also Fig.S5 to S10 for additional fluorescence images).

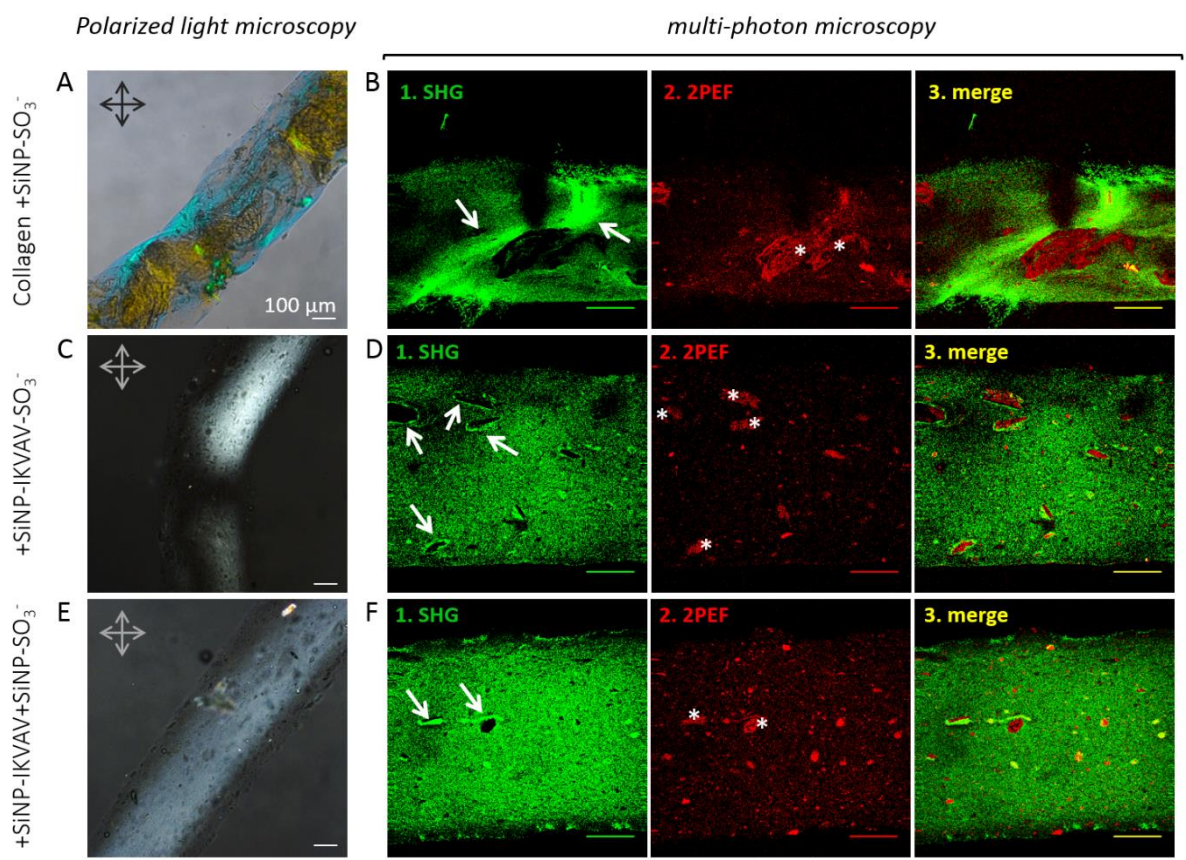

**Fig. 4.** Collagen-SiNP filaments ([Collagen] = 13.5 mg.mL$^{-1}$, [particles] = 45 mg.mL$^{-1}$) observed by (A,C,E) PLM: the uniform birefringence reflected by the brilliance of the material under PLM is indicative of uniaxial alignment (a waveplate was added in (A) to improve contrast, arrows represent cross polarizers); and (B,D,F) multi-photon microscopy combining (1) SHG and (2) 2-PEF with (3) their merge. (A,B) Collagen + SiNP-SO$_3^-$; (C,D) + SiNP-IKVAV-SO$_3^-$; (E,F) + SiNP-IKVAV + SiNP-SO$_3^-$. Scale bars 100 µm.

The pure collagen filament showed a homogeneous birefringence that reflects the alignment of collagen fibrils (Fig. 3A). Concomitantly, a high SHG signal was observed that confirms the tight unidirectional alignment of collagen triple helices within fibrils (Fig.3B1). It is worth noting that the surface of the filament shows the formation of larger structures. This results from the fibrillation process during extrusion that is different at the interface with the solvent and within the filament where solvent diffusion is limited. Such structure was found to be common to all conditions. No fluorescence was observed, in agreement with the absence of SiNPs (Fig.3B2). After incorporation of *nf*-SiNPs, no variation in birefringence was observed (Fig.3C), together with a high SHG signal interrupted by domains with no SHG signals (Fig.3D1). This is attributed to the presence of SiNPs that are aggregated in the filament as observed by 2PEF (Fig.3D2) and confirmed by merging the SHG and 2PEF signals (Fig.3D3). Despite the presence of those *nf*-SiNP aggregates, the collagen organization was not significantly affected, with a homogeneous SHG signal similar to the one of the pure collagen filament. 3D reconstructions (movies) obtained by imaging the filament through its total thickness with steps of 10 µm allow for a clear visualization of collagen organization in 3D (see online Supplementary Materials). On the contrary, the SiNP-IKVAV filaments showed a high homogenous birefringence with a strong SHG signal and a very good dispersion of the particles (Fig.3E,F). This is attributed to favourable interactions between the conjugated IKVAV peptide and the collagen-protein scaffold that prevent particles from aggregation. After incorporation of SiNP-SO$_3^-$, a very different biomaterial is obtained. The collagen-SiNP-SO$_3^-$ filament exhibited a strong birefringence indicative of collagen alignment but with the formation of domains together with local extinction of the birefringence, signing for areas where collagen fibrils are not aligned (Fig.4A). This is also clearly evidenced by SHG: areas of high intensity (white arrows, Fig. 4B1) are separated with dark, lower SHG-intensity areas. The 2PEF image shows the formation of large, ca. 200 µm, SiNP-SO$_3^-$ domains (Fig.4B2, white stars). Merging SHG and 2PEF images shows that the particle domains are surrounded by high-intensity SHG areas (Fig.4B3). Again, this is clearly visible in the corresponding 3D reconstruction (see Supplementary Materials). In those composites, collagen-rich domains with highly aligned fibrils co-exist with SiNP-SO$_3^-$ domains. After incorporation of bifunctional SiNP-IKVAV-SO$_3^-$, heterogeneities in the birefringence were observed under PLM together with local extinction of the birefringence (Fig.4C). This can be attributed to variation in collagen packing throughout the filament. Observations under SHG showed an overall high SHG intensity signing for collagen fibrils, together with small areas with no SHG signal. These areas correspond to the presence of SiNP-IKVAV-SO$_3^-$ aggregates of small size evidenced by 2PEF (white stars, Fig.4D2,3). Interestingly, those areas are surrounded by a bright halo with SHG intensity higher than in the rest of the filament (white arrows, Fig.4D1). This suggests that the presence of SiNP-IKVAV-SO$_3^-$ aggregates increases locally the density or unidirectional alignment of collagen fibrils. As a comparison, filaments were prepared incorporating a mixture of mono-functionalized particles (SiNP-IKVAV + SiNP-SO$_3^-$). In this case, the filaments exhibited a low birefringence by PLM (Fig.4E). SHG imaging showed fibrillar collagen with high SHG intensity. Interestingly, small elongated structures with higher SHG intensity can be observed (white arrows, Fig.4F1). 2PEF imaging reveals the formation of particle aggregates together with a continuous fluorescent signal. This can be attributed to the formation of SiNP-SO$_3^-$ aggregates coexisting with SiNP-IKVAV homogeneously distributed within the filament as previously observed (Fig.4B2 and Fig.3F2 respectively). Merging SHG and 2PEF images indicates that the small elongated structures with high SHG signal co-localize with aggregates, hence with SiNP-SO$_3^-$. Additional imaging using polarization-resolved SHG further confirmed the specific alignment of collagen molecules at the vicinity of the aggregates (Fig.S11)

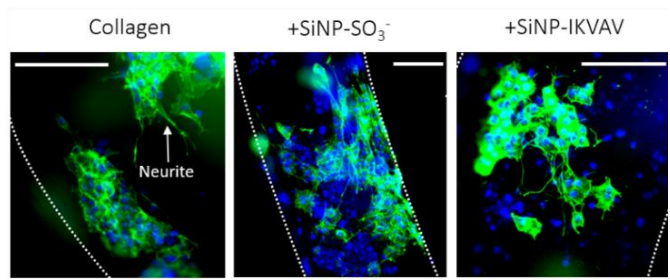

**Fig. 5.** PC12 cells cultured for 10 days on collagen-SiNP filaments observed by fluorescence microscopy. Actin (green: phalloidin) and nucleus (blue: DAPI). The white arrow shows a cell extension called neurite. Scale bars 200 µm. ([Collagen] = 13.5 mg.mL$^{-1}$, [particles] = 45 mg.mL$^{-1}$).

### PC12 cell differentiation

PC12 cells were cultured on collagen-SiNP filaments in presence of FGF1 and heparin. Addition of heparin was essential for differentiation of PC12 cells as it makes up the neurotrophic activity of exogenous FGF1.[50] PC12 adhesion and proliferation were determined by Alamar blue assay and counting the number of cell nuclei. Neuronal differentiation was assessed by quantitative morphological analysis: this included the measurement of the number of extensions per cell, called neurites, having a length larger than the cell body size; the neurite length and alignment along the filament main axis (Fig.5).

Cell number on composite filaments significantly increased in presence of SiNP-IKVAV compared to pure collagen (Fig.6A), showing a beneficial effect of the laminin epitope on cell adhesion after 10 days. Metabolic activities were found to be similar in presence of SiNPs regardless of the surface functionalization (Fig.6B). In terms of number of neurites per cell, no improvement was observed when compared to the pure collagen materials after addition of *nf*-SiNP, SiNP-IKVAV and bifunctional SiNP-IKVAV-SO$_3^-$ (Fig.6C). However, a significant increase was observed after incorporation of SiNP-SO$_3^-$ and SiNP-IKVAV+SiNP-SO$_3^-$. These improved performances were also confirmed when considering neurite length, which significantly increased in presence of SiNP-SO$_3^-$ and SiNP-IKVAV+SiNP-SO$_3^-$ (Fig.6D). We then examined neurite orientation that was defined as "aligned" when the neurite forms an angle inferior to 45° with the axis of the filament. For this parameter, a significant improvement was observed not only in presence of SiNP-SO$_3^-$ and SiNP-IKVAV+SiNP-SO$_3^-$ but also after incorporation of *nf*-SiNPs (Fig.6E).

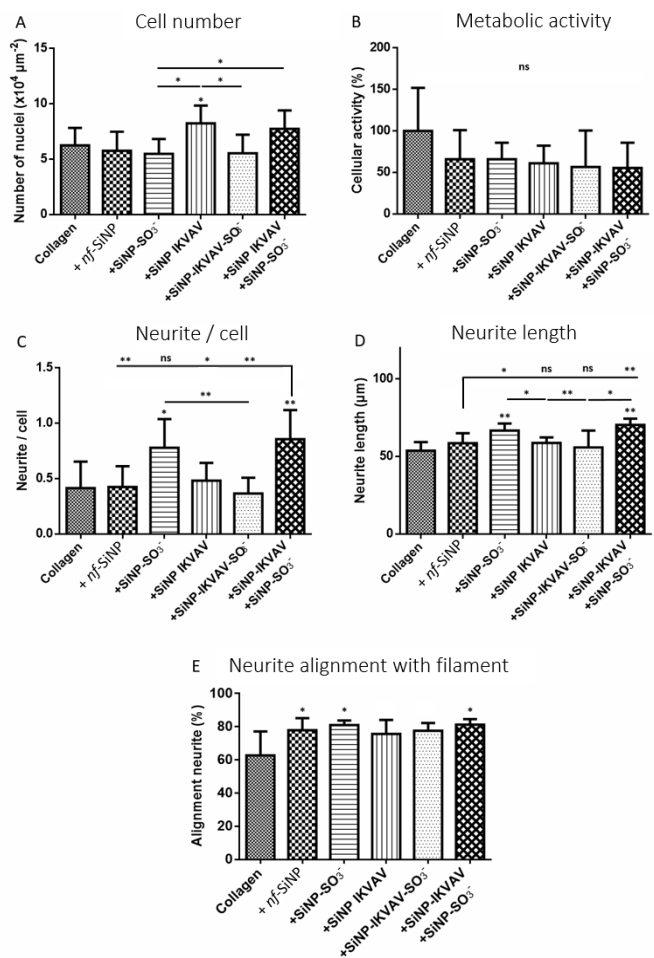

**Fig. 6.** PC12 cells after 10 days of culture on collagen-SiNP filaments ([Collagen] = 13.5 mg.mL$^{-1}$, [SiNP] = 45 mg.mL$^{-1}$). (A) Number of nuclei by unit of filament surface, (B) metabolic activity from Alamar blue assay, (C) number of neurites per cell, (D) neurite length and (E) alignment along the filament. The column represents mean ± standard error of mean (SD) (* $p < 0.05$, ** $p < 0.01$; calculated against the pure collagen filaments, using one-tailed Wilcoxon-Mann-Whitney non parametric test; at least 5 filaments analyzed).

Discussion

Overall, these results show that cell differentiation was significantly improved in presence of SiNP-SO$_3^-$ (either alone or mixed with SiNP-IKVAV), as observed with the increase in neurite number and length (Fig. 6C-D). No such result was obtained with *nf*-SiNP, SiNP-IKVAV or with bifunctional SiNP-IKVAV-SO$_3^-$. We have previously shown that positively charged soluble collagen interacts strongly and specifically through electrostatic interactions with sulfonate-modified SiNPs. This generates composite buildings blocks, where collagen is confined at the particle surface.[29] This is the case here, where SiNPs are mixed with soluble collagen before extrusion in buffer. Then, upon extrusion in buffer, and due to the consecutive pH increase, collagen fibrillogenesis is induced from the particle surface. The formation of those hybrid building blocks (SiNP-SO$_3^-$ / soluble collagen) is an efficient way to trigger collagen fibrillogenesis specifically from the surface of the SiNP-SO$_3^-$. In such case, we have shown that the collagen density decreases with increasing distance from the particle surface, i.e. collagen density is higher at the surface than far from it.[30] At the scale of the biomaterial, this impacts collagen distribution in the heterogeneous filament obtained with SiNP-SO$_3^-$, with the formation of particle-rich domains surrounded by areas with highly aligned collagen observed by PLM, SHG and 2PEF (Fig. 4A,B). This creates a micrometer-scale heterogeneity in collagen organization that cell extensions may detect when sensing their surroundings (Fig. 7A). The resulting modulations of the tissue organization arising at the scale of the cell may be responsible for the increase in neurite number and length. Indeed, growth cone- mediated neuronal elongation is viewed as a result of multiple cytoskeletal dynamics like polymerization of actin and tubulin subunits at the tip of the axon.[51] They are able to sense and generate local stiffness. Our observation can be related to previous works by Sundararaghavan and co-workers who developed a microfluidic chip with controlled gradients of mechanical stiffness in 3D collagen gels, where neurites grow significantly longer down the gradient of stiffness than up the gradient and where no change in cell behavior was observed in absence of gradient.[52]

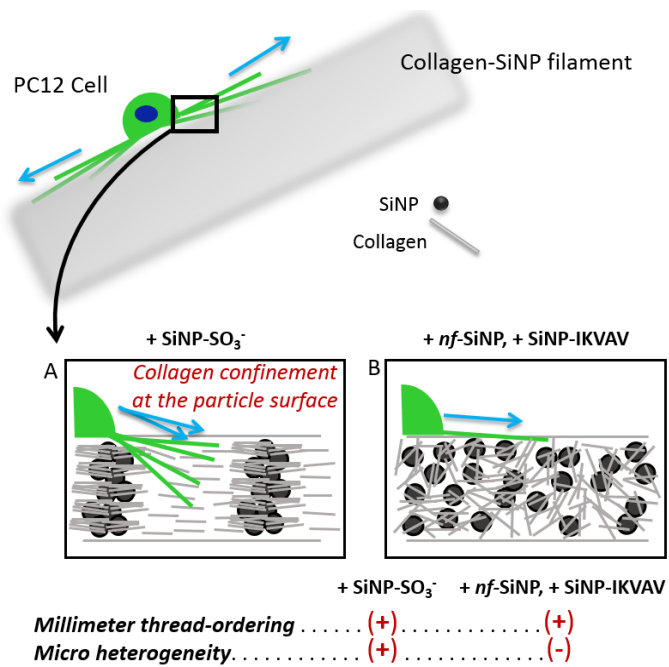

**Fig. 7.** PC12 cells seeded on collagen-SiNP filaments. Scheme of the impact of SiNP surface chemistry and resulting milli- and micrometer-scale ordering on neurite alignment, cell adhesion and differentiation. (A) SiNP-SO$_3^-$; (B) *nf*-SiNP and SiNP-IKVAV.

On the contrary, PLM, SHG and 2PEF observations indicate the formation of small particle aggregates or well dispersed particles for *nf*-SiNPs and SiNP-IKVAV respectively (Fig. 3C-F). This suggests that the material heterogeneity is at a lower -nanometer- scale than the cells so that the filament surface may be detected as a homogeneous substrate of mixed silica/collagen composition (Fig. 7B). Meanwhile, and notwithstanding the micro-scale heterogeneity, neurite alignment with respect to the filament is well preserved in presence of nanoparticles, with a limited impact of their surface chemistry. Neurite alignment is here directed by the millimetre-scale ordering of the filaments that results from the extrusion process.

This allows to define three scales of ordering: (1) the nanostructure of the composite, as defined by the interactions between SiNPs and collagen driving the formation of evenly dispersed aggregates; (2) the micron-scale organization of SiNP-collagen domains that influences cell differentiation, and (3) the millimeter-scale ordering of collagen-based filaments that ensures neurite alignment.

Let's now consider bifunctional biomaterials, i.e. filaments that combine sulfonate groups and IKVAV. The mixture of the two sets of monofunctional SiNP-IKVAV and SiNP-SO$_3^-$, improved neurite number and length to a similar extent as in presence of SiNP-SO$_3^-$ only. Again, SHG images showed the formation of high SHG intensity structures. This indicates that the microstructure of the filament may prevail over the presence of laminin epitope. This is consistent with the absence of improvement of differentiation when only SiNP-IKVAV is present. Thus PC12 cells differentiation seems not to occur when IKVAV is present in our system. This may be attributed to the fact that SiNPs are embedded throughout the entire thickness of the filament (ca. 300 μm in diameter). Because the penetration of PC12s within the filament is limited, a very small amount of laminin epitope is actually interacting with cells. The embedment of SiNP-IKVAV within collagen filaments may render the peptide poorly accessible to the cells. Overall, the concentration of IKVAV laminin epitope on the SiNP surface hence accessible to cells was high enough for improving cell adhesion but not efficient in transducing differentiation.

Noticeably, when bifunctional filaments were prepared by incorporating SiNP-IKVAV-SO$_3^-$ particles, no improvement in neurite number or size was observed compared to pure collagen. This indicates that surface-initiated collagen fibrillogenesis around the particle aggregates is too limited when the two functional groups are on the same particle. This may originate from the partial screening of the sulfonate negatively charged groups by the positive charges of IKVAV, as supported by zeta potential measurements (Fig. 2).

All of those results converge towards the conclusion that the bioactivity of here-described collagen-based biocomposites is dominated by the heterogeneity in collagen density at the microscopic scale, as induced by the addition of SiNP-SO$_3^-$ in soluble collagen before extrusion and fibrillogenesis. The presence of the laminin epitope has a slight beneficial effect on cell adhesion but no impact on differentiation, probably because of its limited accessibility. Very interestingly, this highlights the crucial feature of the microscopic architecture of the biomaterial that in this particular system is more effective than the introduction of a bioactive epitope. Further developments in terms of material processing are now necessary to take advantage of the additional possibility to use silica particles as epitope-presenting components. This includes working with thinner biomaterials or directing SiNPs at the surface of the composite. This would represent an important step forward the engineering of 3D tissue repair scaffolds recapitulating the multi-faceted features of cells microenvironment in ECM.

## Acknowledgements
We thank Gervaise Mosser and Léa Trichet (LCMCP) for fruitful discussions on the fibrillogenesis of collagen filaments upon extrusion, Tristan Baumberger and Olivier Ronsin (INSP) for their help with filament extrusion, Kenta Matsumoto for his help in nanoparticle synthesis and Clothilde Raoux for helping for SHG microscopy. We acknowledge financial support from the French state funds managed by the ANR within the Investissements d'Avenir program under reference ANR-11-IDEX-0004-02, and more specifically within the framework of the Cluster of Excellence MATISSE led by Sorbonne Université.

## Conflicts of interest
There are no conflicts to declare.

1. H.-E. Rosberg, K. S. Carlsson and L. B. Dahlin, *Scand. J. Plast. Reconstr. Surg. Hand. Surg.,* 2005, **39**, 360.
2. G. Stoll and H. W. Müller, *Brain Pathol.,* 1999, **9**, 313.
3. R. V. Bellamkonda, *Biomaterials,* 2006, **27**, 3515.
4. C. Aijie, L. Xuan, L. Huimin, Z. Yanli, K. Yiyuan, L. Yuqing and S. Longquan, *Nanomedicine,* 2018, **13**, 1067.
5. W. Zhu, C. O'Brien, J. R. O'Brien and L. G. Zhang, *Nanomedicine*, 2014, **9**, 859.
6. P. X. Ma, *Adv. Drug Deliv. Rev.*, 2008, **60**, 184.
7. H. R. Hoogenkamp, G.-J. Bakker, L. Wolf, P. Suurs, B. Dunnewind, S. Barbut, P. Friedl, T. H. van Kuppevelt and W. F. Daamen, *Acta Biomater.*, 2015, **12**, 113.
8. E. S. Lai, C. M. Anderson and G. G. Fuller, *Acta Biomater.,* 2011, **7**, 2448.
9. S. Kehoe, X. F. Zhang and D. Boyd, D. *Injury,* 2012, **43**, 553.
10. X. Cao and M. S. Shoichet, *Neuroscience,* 2003, **122**, 381.
11. M.-N Labour, A. Banc, A. Tourrette, F. Cunin, J.-M. Verdier, J.-M. Devoisselle, A. Marcilhac and E. Belamie, *Acta Biomater.,* 2012, **8**, 3302.
12. D. McDonald, C. Cheng, Y. Chen and D. Zochodne, *Neuron Glia Biol.,* 2006, **2**, 139.
13. L. Luckenbill-Edds, *Brain Res. Brain Res. Rev.,* 1997, **23**, 1.
14. M. Jucker, H. K. Kleinman and D. K. Ingram, *J. Neurosci. Res.*, 1991, **28**, 507.
15. E. Agius, Y. Sagot, A. M. Duprat and P. Cochard, *Neuroscience,* 1996, **71**, 773.
16. T. T. Yu and M. S. Shoichet, *Biomaterials,* 2005, **26**, 1507.
17. N. Rangappa, A. Romero, K. D. Nelson, R. C. Eberhart and G. M. Smith, *J. Biomed. Mater. Res.,* 2000, **51**, 625.
18. A. Matsuda, H. Kobayashi, S. Itoh, K. Kataoka and J. Tanaka, *Biomaterials,* 2005, **26**, 2273.
19. G. A. Silva, C. Czeisler, K. L. Niece, E. Beniash, D. A. Harrington, J. A. Kessler and S. I. Stupp, *Science,* 2004, **303**, 1352.
20. E. J. Berns, S. Sur, L. Pan, J. E. Goldberger, S. Suresh, S. Zhang, J. A. Kessler and S. I. Stupp, *Biomaterials,* 2014, **35**, 185.
21. R. A. Que, D. R. Crakes, F. Abdulhadi, C.-H. Niu, N. A. Da Silva and S. W. Wang, *Biotechnol. J.,* 2018, **13**, e1800140.
22. R. A. Que, J. Arulmoli, N. A. Da Silva, L. A. Flanagan and S.-W. Wang, *J. Biomed. Mater. Res. A.,* 2018, **106**, 1363.
23. K. Luo, X. Gao, Y. Gao, Y. Li, M. Deng, J. Tan, J. Gou, C. Liu, C. Dou, Z. Li, Z. Zhang, J. Xu and F. Luo, *Acta Biomater.*, 2019, **85**, 106.
24. M. Guvendiren and J. A. Burdick. *Curr. Opin. Biotechnol.*, 2013, **24**, 841.
25. C. N. Salinas and K. S. Anseth, *J Tissue Eng Regen Med*., 2008, **2**, 296.
26. S. Sur, E. T. Pashuck, M. O. Guler, M. Ito, S. I. Stupp and T. Launey, *Biomaterials,* 2012, **33**, 545.
27. S. Sur, M. O. Guler, M. J. Webber, E. T. Pashuck, M. Ito, S. I. Stupp and T. Launey, *Biomater. Sci.,* 2014, **2**, 903.
28. C. Aimé and T. Coradin, T. *J. Polym. Sci. Part B: Polym. Phys.*, 2012, **50**, 669.
29. C. Aimé, G. Mosser, G. Pembouong, L. Bouteiller and T. Coradin, *Nanoscale,* 2012, **4**, 7127.
30. S. Bancelin, E. Decencière, V. Machairas, C. Albert, T. Coradin, M.-C. Schanne-Klein and C. Aimé, *Soft Matter*, 2014, **10**, 6651.
31. S. Heinemann, T. Coradin and M. F. Desimone, *Biomater. Sci.,* 2013, **1**, 688.
32. S. Jing, D. Jiang, S. Wen, J. Wang and C. Yang, *J. Porous Mater.,* 2014, **21**, 699.
33. S. Heinemann, C. Heinemann, H. Ehrlich, M. Meyer, H. Baltzer, H. Worch and T. Hanke, *Adv. Eng. Mater.,* 2007, **9**, 1061.
34. Z. Zhang, Y. Liu, X. Zhu, L. Wei, J. Zhu, K. Shi, G. Wang and L. Pan, *J. Vet. Sci.,* 2018, **19**, 512.
35. R. Hu, Q. Cao, Z. Sun, J. Chen, Q. Zheng and F. Xiao, *Int. J. Mol. Med.,* 2018, **41**, 195.
36. N. F. Eskici, S. Erdem-Ozdamar and D. Dayangac-Erden, *Cell. Mol. Biol. Lett.,* 2018, **23**, 5.
37. L. Zhang, H. Dong, Y. Si, N. Wu, H. Cao, B. Mei and B. Meng, *Neurobiol. Aging,* 2019, **73**, 41.
38. R. E. Rydel and L. A. Greene, *J. Neurosci.*, 1987, **7**, 3639.
39. P. A. Walicke, *J. Neurosci.*, 1988, **8**, 2618.
40. A. Rodriguez-Enfedaque, S. Bouleau, M. Laurent, Y. Courtois, B. Mignotte, J.-L. Vayssière and F. Renaud, *Biochim. Biophys. Acta,* 2009, **1793**, 1719.
41. S. Bouleau, I. Pârvu-Ferecatu, A. Rodriguez-Enfedaque, V. Rincheval, H. Grimal, B. Mignotte, J.-L. Vayssiere and F. Renaud, *Apoptosis,* 2007, **12**, 1377.
42. H.-C. Ni, T.-C. Tseng, J.-R. Chen, S.-h. Hsu and I.-M. Chiu, *Biofabrication*, 2013, **5**, 035010.
43. Y.-C. Hsu, S.-L. Chen, D.-Y. Wang, I.-M. Chiu, *Biomed. J.* 2013, 36, 98.
44. L. Picaut, L. Trichet, O. Ronsin, B. Haye, I. Génois, T. Baumberger and G. Mosser, *Biomed. Phys. Eng. Express* 2018, **4**, 035008.
45. N.-J. Jan, J. L. Grimm, H. Tran, K. L. Lathrop, G. Wollstein, R. A. Bilonick, H. Ishikawa, L. Kagemann, J. S. Schuman and I. A. Sigal, *Biomed. Opt. Express,* 2015, **6**, 4705.
46. S. Bancelin, A. Nazac, B. H. Ibrahim, P. Dokládal, E. Decencière, B. Teig, H. Haddad, H. Fernandez, M.-C. Schanne-Klein and A. De Martino. *Opt Express.,* 2014, **22**, 22561.
47. S. Bancelin, C. Aimé, T. Coradin and M.-C. Schanne-Klein, *Biomed. Opt. Express,* 2012, **6**, 1446.
48. S. Bancelin, C. Aimé, I. Gusachenko, L. Kowalczuk, G. Latour, T. Coradin and M.-C. Schanne-Klein, *Nat. Commun.*, 2014, **5**, 4920.
49. X. Chen, O. Nadiarynkh, S. Plotnikov and P. J. Campagnola, *Nat. Protoc.,* 2012, **7**, 654.
50. F. Renaud, S. Desset, L. Oliver, G. Gimenez-Gallego, E. Van Obberghen, Y. Courtois and M. Laurent, *J. Biol. Chem.*, 1996, **271**, 2801.
51. D. M. Suter and K. E. Miller, *Prog. Neurobiol.*, 2011, **94**, 91.
52. H. G. Sundararaghavan, G. A. Monteiro, B. L. Firestein and D. I. Shreiber, *Biotechnol. Bioeng.* 2009, **102**, 632.